\documentclass[12pt,letterpaper]{article}
\usepackage[T1]{fontenc}
\usepackage[utf8]{inputenc}
\usepackage[greek,english]{babel}
\usepackage[font=small, labelfont=bf]{caption}
\usepackage{graphicx}
\usepackage{amsmath}
\usepackage{amssymb}
\usepackage{amsthm}
\usepackage{longtable}
\usepackage{verbatim}
\usepackage{cite}
\usepackage{enumerate}
\usepackage{hyperref}
\usepackage{subfig}  
\usepackage[usenames,dvipsnames]{xcolor}
\usepackage{bbding}
\usepackage{MnSymbol}
\usepackage{bbold}

\newcommand{\ZZ}{{\mathbb{Z}}}

\newcommand{\Uone}{{U(1)}}

\newcommand{\UE}{{E}}

\newcommand{\mf}{\mathfrak}
\newcommand{\mfg}{{\mathfrak{g}}}

\theoremstyle{definition}

\usepackage{titlesec}

\def\be{\begin{equation}}
\def\ee{\end{equation}}

\begin{document}
	\newcommand{\main}{.}
\begin{titlepage}

\setcounter{page}{1} \baselineskip=15.5pt \thispagestyle{empty}

\bigskip\

\bigskip\

\vspace{2cm}
\begin{center}
{\LARGE \bfseries Nonabelian Lattice Weak Gravity Conjecture \\ \vspace{.2cm} and Monopole Confinement }

 \end{center}
\vspace{0.5cm}

\begin{center}
{\fontsize{14}{30}\selectfont Matthew Reece$^a$ and Tom Rudelius$^b$}
\end{center}

\begin{center}

\vspace{0.25 cm}
\textsl{$^a$Department of Physics, Harvard University, Cambridge, MA, 02138, USA}\\
				\textsl{$^b$Department of Mathematical Sciences, Durham University, Durham, DH1 3LE, UK}\\

\end{center}

\vspace{1cm}

\noindent 
Within the known landscape of quantum gravity, most theories satisfy the Lattice Weak Gravity Conjecture (LWGC), which requires a superextremal particle at every site in the electric charge lattice $\Gamma$. However, counterexamples to the LWGC exist, and it was recently hypothesized that such counterexamples necessarily feature fractionally charged confined monopoles. In this work, we verify this hypothesis in toroidal orbifold compactifications of the heterotic string, which notably feature LWGC violation in both the abelian and nonabelian gauge sectors. In all the cases we consider, there exists a discrete subgroup of the center of the gauge group $K \subseteq Z(G)$ such that superextremal particles exist at every site in the charge lattice of the quotient group $G/K$, while (confined) monopoles exist at all sites in the magnetic charge lattice of $G/K$. This suggests that LWGC violation cannot occur for gauge groups with trivial centers, and more generally the degree of LWGC violation in a nonabelian gauge theory is bounded in terms of the maximal order of the center.

 \vspace{1.1cm}

\bigskip
\noindent\today

\end{titlepage}
\setcounter{tocdepth}{2}
\tableofcontents

\section{Introduction}

In the quest to bridge the gap between string theory and experimental physics, the Weak Gravity Conjecture (WGC) \cite{Arkanihamed:2006dz} has emerged as one of the key stepping stones. The original, vanilla version of the conjecture holds that an in any $U(1)$ gauge theory, there must exist a ``superextremal'' particle whose charge-to-mass ratio is greater than or equal to that of a large extremal black hole:
\be
\frac{|q|}{m} \geq \frac{|Q|}{M}\Big|_{\textrm {large ext BH}} \equiv Z_{\rm ext}\,.
\label{WGCbound}
\ee
Even in the original paper on the WGC, it was observed that this mild version of the WGC can be strengthened in various ways. Since then, many different variants of the WGC have been proposed, which lead to different constraints on particle physics, cosmology, mathematics, and more. Determining which of these versions are correct remains an important outstanding problem.

In this work, we will focus our attention on tower/lattice variants of the WGC, which were initially proposed in \cite{Heidenreich:2015nta, Heidenreich:2016aqi, Andriolo:2018lvp}. The strongest of these is the Lattice WGC (LWGC), which holds that \emph{every} site in the charge lattice of a $U(1)$ gauge
theory in quantum gravity must have a particle that satisfies the Weak Gravity bound \eqref{WGCbound}. In theories with multiple $U(1)$s, the LWGC compares the charge-to-mass ratio of
each particle to that of a large extremal black hole in the same direction in the charge
lattice.

Counterexamples to the LWGC were discovered in \cite{Heidenreich:2016aqi}, based on earlier ideas in \cite{Arkanihamed:2006dz}. Consequently, \cite{Heidenreich:2016aqi} proposed a slightly weakened version of the LWGC known as the Sublattice WGC (sLWGC), which requires superextremal particles at a finite-index sublattice $\Gamma_{\rm ext}$ of the electric charge lattice of the theory, $\Gamma$. Said differently, the sLWGC implies that there exists an integer $k$ such that for any $\vec q \in \Gamma$, there exists a superextremal particle of charge $k \vec q$. The smallest such $k$ is referred to as the \emph{coarseness} of the sublattice $\Gamma_{\rm ext}$.

The sLWGC can be extended to nonabelian gauge theories as well. The simplest way to do this is simply to fix a Cartan subgroup $U(1)^n \subseteq G$ of the gauge group $G$ and apply the sLWGC to the theory with the Cartan subgroup. With this, the charge lattice of the nonabelian sector of the gauge theory is given simply by its weight lattice. A stronger variant of the sLWGC for nonabelian gauge groups was introduced in \cite{Heidenreich:2017sim}, which holds that for every dominant weight $\vec w \in \Gamma_{\rm ext}$, there exists a superextremal particle transforming in the irreducible representation with highest weight $\vec{w}$. These two variants become equivalent in the absence of very superextremal particles of large charge ($|\vec Q|, |\vec Q|/m \gg 1$ in Planck units). Within the class of heterotic string theories studied in the present work, such particles do not arise,\footnote{The mass of a heterotic string state $m$ is bounded below in terms of its charge $\vec P$ as $\frac{1}{4} m^2 \alpha' \geq \frac{P^2}{2}-1$. This ``$-1$ offset'' is too small to allow for violations of the stronger version of the nonabelian sLWGC without also violating the weaker version.} so these two versions of the nonabelian sLWGC are equivalent.

The failure of the LWGC raises a number of important questions: what fundamental principle determines whether or not a theory will satisfy the LWGC?
If it can be violated, can the sLWGC be violated as well? If not, what is the maximal allowed value of the coarseness, $k$? 

The recent work \cite{Etheredge:2025rkn} gave partial answers to these questions, providing compelling evidence that the hallmark of LWGC violation, in theories that satisfy the sLWGC, is the presence of fractionally charged confined monopoles. Such monopoles are not genuine monopoles in the theory, as they violate Dirac quantization. Instead, they are confined via flux tubes (that is, they are attached to physical Dirac strings). The work \cite{Etheredge:2025rkn} showed that in multiple classes of gravitional EFTs (including several UV complete examples), fractionally charged confined monopoles are ``dual'' to LWGC violation in the sense that 
\be
\Gamma_{\rm ext} \supseteq \tilde \Gamma_{\rm con}^\vee \,,
\label{extduality}
\ee
where $\Gamma_{\rm ext} \subseteq \Gamma$ is a sublattice of superextremal particles, $\tilde \Gamma_{\rm con} \supseteq \tilde \Gamma$ is the superlattice of magnetic monopole charges including the confined fractional charges, $\Gamma$ ($\tilde \Gamma$) is the electric (magnetic) charge lattice, and $^\vee$ denotes the lattice dual. This immediately implies that the coarseness of the superextremal sublattice $\Gamma_{\rm ext} \subseteq \Gamma$ is bounded above by the coarseness of the magnetic monopole lattice $\tilde \Gamma \subseteq \tilde \Gamma_{\rm con}$, and the sLWGC is satisfied even though LWGC may be violated.\footnote{There are theories in which the sLWGC is not known to be true (or false)~\cite{Lee:2019tst,Cota:2022yjw,Cota:2022maf,Casas:2024ttx}. For a proposed alternative conjecture, see~\cite{FierroCota:2023bsp}.}

The examples studied in \cite{Etheredge:2025rkn} featured limits in which charged particles outside the superextremal sublattice $\Gamma_{\rm ext}$ became heavy and the flux tubes between confined monopoles became light, as measured in Planck units. This limit can be thought of as a dynamical change of the gauge group from $G$ to $G/K$, and accordingly the electric charge lattice shrinks to a sublattice $\Gamma_{\rm ext}$ while the magnetic charge lattice expands to a superlattice $\tilde \Gamma_{\rm con}$ \cite{Aharony:2013hda, Gaiotto:2014kfa}. Said differently, when particles outside $\Gamma_{\rm ext}$ become infinitely heavy and exit the spectrum, leaving behind a 1-form symmetry $K \cong \Gamma / \Gamma_{\rm ext}$. Simultaneously, the fractionally charged monopoles deconfine and become genuine monopoles of the theory, indicating that the emergent symmetry $K$ is in fact a 1-form \emph{gauge} symmetry rather than a global symmetry. 

The examples considered in \cite{Etheredge:2025rkn} all dealt with LWGC violation in $U(1)$ gauge theories. In this work, we expand the study of LWGC violation and fractional monopole confinement to the nonabelian sector, focusing especially on toroidal orbifold compactifications of the heterotic string with Wilson lines. We find compelling evidence that the relation \eqref{extduality} holds in this context as well, providing further evidence for the connection between LWGC violation and monopole confinement. In addition, we find that in all of these examples, the superextremal sublattice $\Gamma_{\rm ext}$ can be identified with the charge/weight lattice of a discrete quotient of the gauge group, $G_{\rm ext} = G/K$, with $K \subseteq Z(G)$ a subgroup of the center $Z(G)$, so that $K \cong \Gamma / \Gamma_{\rm ext}$. The coarseness of $\Gamma_{\rm ext}$ can thus be identified with the maximal order of the elements of $K$. 

These results immediately suggest an important corollary: if $Z(G)$ is trivial, the LWGC must be satisfied. Indeed, we will verify that the LWGC is satisfied (and in fact saturated) in the 9d theory with $E_8$ gauge group constructed by circle compactification of $[E_8 \times E_8] \rtimes \mathbb{Z}_2$ gauge theory with a $\mathbb{Z}_2$ Wilson line turned on. Turning the logic around, one might say that the connection between LWGC violation and fractional monopole confinement uniquely fixes the famous ``$-1$ offset'' in the 10d heterotic string spectrum, $\frac{1}{4} m^2 \alpha'= \frac{\vec q^2+ \vec p_L^2}{2}-1$.

One difference between our examples and those studied in \cite{Etheredge:2025rkn} is that our examples do not seem to feature moduli space limits in which the discrete group $K$ becomes an emergent symmetry. Indeed, for any finite value of the moduli, LWGC violation occurs only at a finite number of sites in the charge lattice. It is noteworthy that a discrete group $K$ still plays a starring role in the story of LWGC violation in these examples, even if its physical interpretation is less clear.

The remainder of this paper is organized as follows: In \S\ref{s.REVIEW}, we review magnetic monopoles for theories with nonabelian gauge groups, and we review the construction of confined monopoles for theories with discrete Wilson lines introduced in \cite{Etheredge:2025rkn}. In \S\ref{s.9dHET}, we show that 9d heterotic string theory with a $\mathbb{Z}_2$ Wilson line saturates the LWGC. In \S\ref{s.T4Z3ex}, we explore monopole confinement of a heterotic orbifold model with LWGC violation and verify \eqref{extduality} under reasonable assumptions. In \S\ref{s.T4Z3ex2}, we consider another such orbifold, focusing particularly on questions of monopole stability. Finally, in \S\ref{s.DISC} we conclude with a discussion of our results and a list of open questions.

\section{Review: Magnetic Monopoles}\label{s.REVIEW}

In this section, we review the classification of monopoles in nonabelian gauge theory introduced in \cite{Lubkin:1963zz,Goddard:1976qe,Kapustin:2005py}. We also review the construction of fractionally charged confined magnetic monopoles in compactifications with Wilson lines described in \cite{Etheredge:2025rkn}.

\subsection{Magnetic monopoles in nonabelian gauge theories}\label{ss.NAREVIEW}

There are two different classifications of magnetic monopoles that are of interest for general gauge theories: a coarse topological classification~\cite{Lubkin:1963zz}, and a refined algebraic classification that allows for a lattice of charges~\cite{Goddard:1976qe}. Here we review the latter from the modern viewpoint of~\cite{Kapustin:2005py}, which exploits UV conformal symmetry to provide a precise characterization. We restrict our attention to four dimensions for the sake of exposition, but the extension to higher dimensions is straightforward (the monopole worldvolume simply acquires $d - 4$ additional spectator dimensions).

Let $G$ be a semisimple Lie group. We consider an 't~Hooft line operator in $G$ gauge theory sitting at $r=0$ in $H^2 \times S^2$, with metric
\be
ds^2 = \frac{dt^2 + dr^2}{r^2} + d \Omega_{2}^2\,.
\ee
The $SL(2, \mathbb{R}) \times SO(3)$ invariance of this space forces the nonabelian field strength to take the form
\be
F = \frac{B}{2} \text{vol}(S^2)+ O(1)\,,
\ee
where $B$ is a covariantly-constant section of the adjoint bundle of $S^2$. As shown in \cite{Goddard:1976qe}, this implies a quantization law for $B$:
\be
\exp(2 \pi i B) = \mathbb{1}_G\,.
\label{Bquant}
\ee
By an appropriate gauge transformation, we can choose $B$ to Lie in a Cartan subalgebra $\mathfrak{t} \subset \mathfrak{g}$.

Following \cite{Kapustin:2005py}, let us first suppose that $G$ has trivial center; that is,
\be
G = \tilde G / Z(\tilde G)\,,
\ee
for $\tilde G$ the simply connected gauge group with Lie algebra $\mathfrak{g}$. Then, the quantization condition \eqref{Bquant} is equivalent to the constraint
\be
\vec{\alpha}(B) \in \mathbb{Z}
\label{alphaconst}
\ee
for all roots $\vec{\alpha}$ of the Lie algebra $\mathfrak{g}$. 

The Langlands dual $^L{\mathfrak{g}}$ of $\mathfrak{g}$ is defined by setting the roots of $\mathfrak{g}$ to be the co-roots of $^L{\mathfrak{g}}$ (and vice versa), while the Cartan torus of $\mfg$ is set equal to the dual of the Cartan torus of $^L\mfg$, $\mathfrak{t} = {^L\mf t}^*$. The Weyl groups of $\mfg$ and $^L\mfg$ are the same, and when $\mfg$ is simply laced, the roots and co-roots are the same, so $\mfg$ is isomorphic to $^L\mfg$.

With this, the requirement \eqref{alphaconst} is equivalent to the condition that $B$ is an element of ${^L\mf t}^*$ that is integer valued on the coroots of $^L{\mfg}$, which is equivalent to the statement that $B$ is a weight of $^L{\mfg}$. $B$ is defined only up to the action of the Weyl group; the quantization condition \eqref{Bquant} is invariant under this action. Thus, we conclude that magnetic monopoles of a semisimple Lie group $G$ with trivial center are classified by weights $w \in \Lambda_{mw}$ of the Langlands dual algebra $^L{\mfg}$ modulo the action of its Weyl group. Equivalently, magnetic monopoles are classified by irreducible representations of $^L\mfg$.

Not all of these monopoles will be stable, however. Stable monopoles are distinguished by nontrivial 't Hooft magnetic fluxes, and they are associated with points in the weight lattice modulo the root lattice of $^L{\mfg}$ (up to Weyl reflections). Equivalently, stable monopoles are classified by elements of $\pi_1(G)$~\cite{Lubkin:1963zz}. Unstable monopoles can decay to stable ones by radiating gauge bosons~\cite{Brandt:1979kk,Coleman:1982cx}.

Let us now drop the assumption that $G$ is centerless. In this case, some weights of $^L\mfg$ are projected out of the magnetic charge lattice $\Lambda_{mw}$, leaving a sublattice $\tilde \Gamma$, which can be identified with the weight lattice of the Langlands dual group $^L G$ with Lie algebra $^L\mfg$. This sublattice satisfies $Z(G) = \Lambda_{mw} / \tilde \Gamma$ and $\Gamma / \Lambda_{cr} = \pi_1(G)$, for $\Lambda_{cr}$ the co-root lattice of $\mfg$.

In this paper, we will be interested in fractionally charged confined monopoles. These monopoles violate Dirac quantization, and thus they are associated with elements of ${^L\mf t}^*$ (up to Weyl reflection) that lie outside the lattice $\tilde \Gamma$. Such a monopole is attached to a physical Dirac string, i.e., it is confined.

\subsubsection{Example: $SU(n)/\mathbb{Z}_k$}

To illustrate, let us first consider the simple example of $G=PSU(n) = SU(n)/\mathbb{Z}_n$. The Langlands dual group is given by $^LG = SU(n)$, so monopoles are classified by irreducible representations of $SU(n)$.

Not all of these representations are stable, however. Stable representations are specified by nontrivial points in the weight lattice of $SU(n)$ modulo the root lattice and the action of the Weyl group. This means, for instance, that the adjoint representation corresponds to an unstable monopole. Equivalently, stable monopoles are characterized by nontrivial elements of $\pi_1(SU(n)/\mathbb{Z}_n) \cong \mathbb{Z}_n$. Here, the fundamental representation $\bf{n}$ of $^LG = SU(n)$ corresponds to charge $1 \mod n$ and the antifundamental representation $\overline{\bf{n}}$ corresponds to charge $-1 \mod n$.

For comparison, let us consider the case $G=SU(n)$. The Langlands dual is then given by $^LG = PSU(n) = SU(n)/\mathbb{Z}_n$, whose weight lattice is equal to its root lattice. This means that monopoles of $SU(n)$ are labeled by irreducible representations of $PSU(n)$, but none of them are stable. Relatedly, $\pi_1(G)$ is trivial, as $SU(n)$ is simply connected. Any monopoles with charges outside the weight lattice of $^LG = PSU(n)$ are fractionally charged and must therefore be confined.

Finally, let us consider the intermediate case of $G=SU(n)/\mathbb{Z}_k$, where $k$ is a positive integer that divides $n$. Let $p=n/k$. Then, monopoles are characterized by irreducible representations of the Langlands dual, which is given by $G=SU(n)/\mathbb{Z}_p$. Stable monopoles are characterized by $\pi_1(^LG) = \mathbb{Z}_p$.

\subsection{Confined monopoles in compactifications with Wilson lines}\label{ss.confinedmonopoles}

In this subsection, we review the construction of confined monopoles in compactifications with Wilson lines used in \cite{Etheredge:2025rkn}. We primarily focus our attention on the case of $\mathbb{Z}_p$ gauge theory compactified on $S^1$, though we comment on generalizations to more sophisticated compactifications with Wilson lines.

To begin, consider $\mathbb{Z}_p$ gauge theory coupled to gravity compactified from $D$ dimensions to $d=D-1$ dimensions on $S^1$ \emph{without} a Wilson line turned on. The resulting theory in $d$ dimensions has gauge group $\mathbb{Z}_p \times U(1)_{\rm KK}$, where $U(1)_{\rm KK}$ is the Kaluza-Klein $U(1)$ associated with the graviphoton. The Kaluza-Klein modes charged under $U(1)_{\rm KK}$ saturate the LWGC at tree level, and the magnetic monopoles are given by the product of a Taub-NUT geometry and $\mathbb{R}^{d-4}$ \cite{Gross:1983hb, Sorkin:1983ns}. The (charge 1) Taub-NUT geometry consists of an $S^1$ fibration over $\mathbb{R}^3$ in which the $S^1$ fiber degenerates at the core of the monopole, yet the total space remains smooth. Due to this degeneration, the Taub-NUT space is simply connected, $\pi_1(\textrm{TN}) = \{1 \}$.

Now, suppose we turn on a Wilson line for the $\mathbb{Z}_p$ gauge field around the circle. As discussed in \S2 of \cite{Etheredge:2025rkn}, this breaks the gauge group of the resulting $d$-dimensional theory from $\mathbb{Z}_p \times U(1)_{\rm KK}$ to a ``diagonal'' $U(1)$. The $n$th Kaluza-Klein mode of a charge $q$ particle under $\mathbb{Z}_p$ will carry charge
$n + \frac{q}{p}$ under this diagonal $U(1)$. If there are no massless particles with charge $q \neq 0$ under $\mathbb{Z}_p$ in $D$ dimensions, the LWGC will be violated after reduction, and the sLWGC will be satisfied with coarseness $p$.

On the magnetic side of the story, there is a puzzle: KK monopoles come from Taub-NUT geometries with trivial $\pi_1(\textrm{TN})$, which means there is no 1-cycle on which to wrap the Wilson line. Naively, this seems to suggest that there are no KK monopoles at all, in violation of the Completeness Hypothesis \cite{polchinski:2003bq}.

\begin{figure}
    \centering
\includegraphics[width=0.6\linewidth]{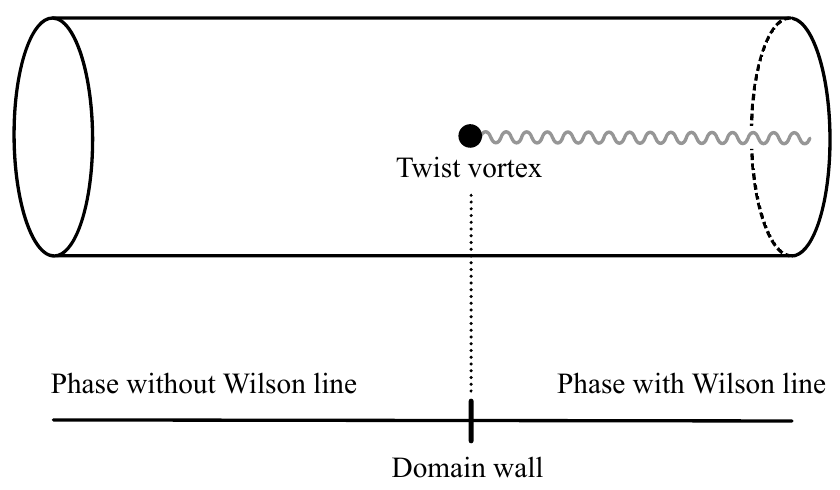}
    \caption{An unwrapped twist vortex acts as a domain wall after circle reduction. Here, the wavy line indicates an unphysical branch cut, which plays a role similar to that of the Dirac string. Figure adapted from \cite{Etheredge:2025rkn}.}
    \label{fig:TVDW}
\end{figure}

The resolution to this puzzle involves twist vortices and confined monopoles. In $D$ dimensions, the Completeness Hypothesis requires the existence of twist vortices---codimension-2 objects characterized by a nontrivial $\mathbb{Z}_p$ holonomy around their worldvolume \cite{Heidenreich:2021tna}. Compactifying to $d$ dimensions, an unwrapped twist vortex acts as a domain wall between one phase in which the Wilson line is turned on and one in which it is turned off, as shown in Figure \ref{fig:TVDW}.

\begin{figure}
    \centering
\includegraphics[width=0.6\linewidth]{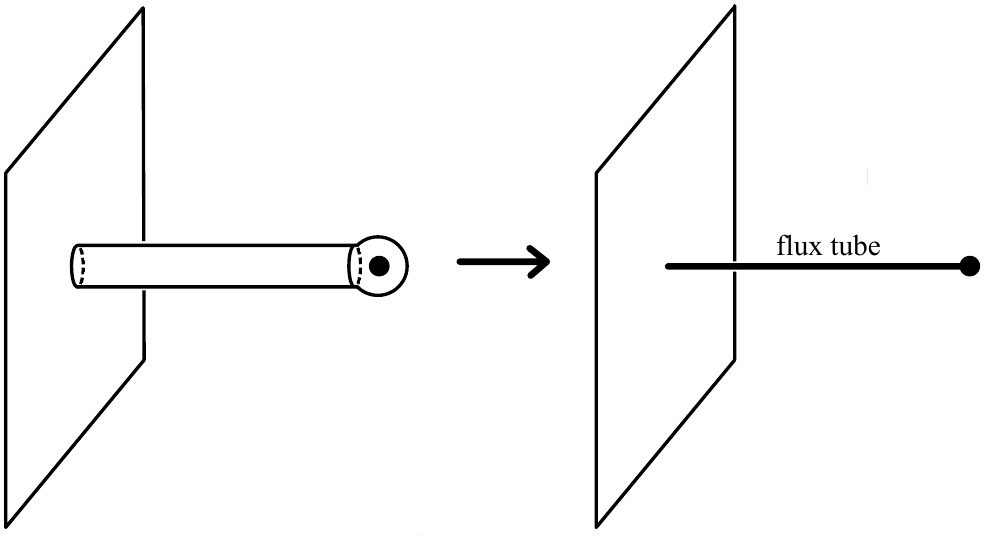}
    \caption{A magnetic monopole in a domain without a Wilson line may be pushed into the domain with the Wilson line by deforming the domain wall between them. Collapsing the tube surrounding the monopole produces a codimension-2 flux tube, hence the monopole is confined in the theory with the Wilson line. Figure adapted from \cite{Etheredge:2025rkn}.}
    \label{fig:DWtube}
\end{figure}

This enables the following construction: start with a (genuine) KK monopole of charge 1 in the vacuum without a Wilson line. Next, allow the monopole to move into a domain with the Wilson line turned on, deforming the domain wall as shown in Figure \ref{fig:DWtube} (left) so as to create a narrow tube in which the monopole resides. Finally, collapse this tube to form a codimension-2 flux tube, as in Figure \ref{fig:DWtube} (right). This flux tube can be constructed by wrapping the twist vortex around the compactification circle.

\begin{figure}
    \centering
\includegraphics[width=0.6\linewidth]{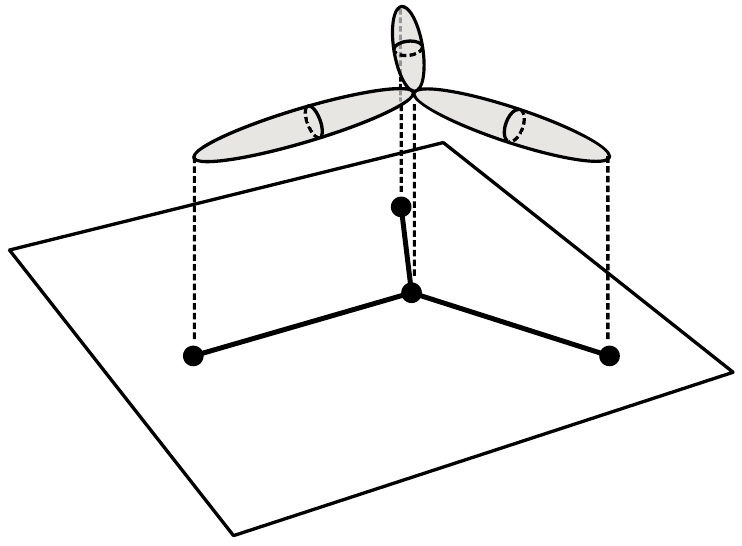}
    \caption{A collection of $p$ confined monopoles may join to form an unconfined monopole of charge $p$. Figure adapted from \cite{Etheredge:2025rkn}.}
    \label{fig:ptubes}
\end{figure}

The result is a confined monopole of magnetic charge 1. $p$ of these confined monopoles may coallesce to form a charge $p$ monopole; since $p \equiv 0 \mod p$, the resulting flux tube is trivial and may be removed, leaving behind a genuine, unconfined, charge $p$ monopole, as shown in Figure \ref{fig:ptubes}. In contrast to the standard case of a confined $U(1)$ gauge group, not {\em all} magnetic flux is collimated into flux tubes. Magnetic field lines will spread out from a charge 1 magnetic monopole, as they would from an unconfined monopole, with compensating magnetic flux carried away by the flux tube (see Figure~18 of~\cite{Etheredge:2025rkn}).

In summary, in the vacuum with the Wilson line turned on, the electric charge lattice is given by $\Gamma = \frac{1}{p} \mathbb{Z}$, while the (genuine) magnetic charge lattice is given by its dual, $\tilde \Gamma = p \mathbb{Z}$.
The sLWGC-satisfying superextremal sublattice consists of integral electric charges, $\Gamma_{\rm ext} = \mathbb{Z}$, while the lattice of fractionally charged magnetic monopoles is given by $\tilde \Gamma_{\rm con} = \mathbb{Z}$. Thus, the sublattice of superextremal particles is dual to the superlattice of confined monopoles, $\Gamma_{\rm ext} = \tilde \Gamma_{\rm con}^\vee$, and \eqref{extduality} is saturated.

More generally, we expect this construction of confined monopoles to work whenever two vacua---one with a discrete Wilson line, one without---are separated by a domain wall. Furthermore, the Cobordism Conjecture of \cite{McNamara:2019rup} suggests that there must exist a finite-tension domain wall between any two such vacua. In the remainder of this work, we will assume that this is always possible, so genuine monopoles in one vacuum can be turned into monopoles of another vacuum that differs by a choice of discrete Wilson line. If these monopoles do not obey Dirac quantization in the latter vacuum, they will be confined by flux tubes.

\section{Example 1: 9d Heterotic String Theory with a $\mathbb{Z}_2$ Wilson Line}\label{s.9dHET}

In this section, we investigate the spectrum of charged particles in $[E_8 \times E_8] \rtimes \mathbb{Z}_2$ gauge theory with a $\mathbb{Z}
_2$ Wilson line turned on. This construction closely mirrors that of \cite{Etheredge:2025rkn} (see also \cite{Heidenreich:2016aqi, Montero:2022vva}), which studied (UV-complete) compactifications of $[U(1) \times U(1)] \rtimes \mathbb{Z}_2$ gauge theory with a $\mathbb{Z}_2$ Wilson line and found LWGC violation. In contrast, we will see that the the LWGC is satisfied in the heterotic case at hand.

\subsection{Reminder: the 10d heterotic theory}

We begin with a review of 10d heterotic string theory with gauge group $[\UE_8 \times \UE_8] \rtimes \ZZ_2$. The charge lattice $\Gamma_8$ of $\UE_8$ consists of $8$-dimensional vectors ${\vec q}$ which are made up of either all integers or all (odd) half-integers, summing to an even number, i.e.:
\begin{align}
\Gamma_8 &= \Gamma_e \cup \Gamma_o, \\
\Gamma_e &= \{ (q_1, \ldots q_8) \, | \, q_i \in \mathbb{Z}, \, \sum_i q_i \in 2 \mathbb{Z} \}, \\
\Gamma_o &= \{ (q_1, \ldots q_8) \, | \, q_i - \frac{1}{2} \in \mathbb{Z}, \, \sum_i q_i \in 2 \mathbb{Z} \}.
\end{align}
In the heterotic string theory, we have states with charges $({\vec q}_1, {\vec q}_2)$ under the two $\UE_8$ factors, with mass determined by
\begin{equation} \label{eq:10dspectrum}
m^2 = \frac{2}{\alpha'} \left( | {\vec q}_1 |^2 + | {\vec q}_2 |^2 - 2 \right).
\end{equation}
These states all satisfy the WGC in 10d, which requires that~\cite{Heidenreich:2015nta}
\begin{equation}
m^2 \leq \frac{g_{10}^2}{\kappa_{10}^2} \left( | {\vec q}_1 |^2 + | {\vec q}_2 |^2  \right).
\end{equation}
We have $g_{10}^2 / \kappa_{10}^2 = 2 / \alpha'$, so the spectrum asymptotically saturates the WGC at large $|{\vec q}|$ but satisfies it with room to spare at small charge.

Because superextremal states of all possible charges $({\vec q}_1, {\vec q}_2)$ exist in the spectrum, we say that the 10d theory respects the Lattice Weak Gravity Criterion.

\subsection{Compactifying to 9d with a twist}

Now we compactify the theory to 9 dimensions on a circle of radius $R$ with a nontrivial $\ZZ_2$ Wilson line (a compactification known as the 9d CHL string~\cite{Chaudhuri:1995fk,Chaudhuri:1995bf,Font:2021uyw}). This is all closely analogous to the case of a gauge group $[\Uone \times \Uone] \rtimes \ZZ_2$ discussed in~\cite{Etheredge:2025rkn}. Only the diagonal $\UE_8$ gauge symmetry survives in the 9d theory. Under the dimensional reduction, a state of mass $m$ and charges $({\vec q}_1, {\vec q}_2)$ becomes a state of mass $m$ and charge 
\begin{equation} \label{eq:chargeadd}
{\vec q} = {\vec q}_1 + {\vec q}_2.
\end{equation}
When matching 10d to 9d, we have
\begin{equation} \label{eq:10dto9d}
\frac{1}{g_9^2} = \frac{4 \pi R}{g_{10}^2}, \quad \frac{1}{\kappa_9^2} = \frac{2 \pi R}{\kappa_{10}^2}.
\end{equation}
The factor of $4\pi R$ rather than $2\pi R$ when matching the gauge couplings arises because both of the 10d $\UE_8$ factors contribute to the kinetic term of the diagonal $\UE_8$ in 9d. As a result, the WGC criterion for a state of charge $\vec q$ in 9d becomes:
\begin{equation} \label{eq:9dWGC}
m^2 \leq \frac{g_9^2}{\kappa_9^2} |{\vec q}|^2 = \frac{1}{\alpha'} |{\vec q}|^2.
\end{equation} 
We would like to know if, for a given ${\vec q}$, we can find a 10d state of charge $({\vec q}_1, {\vec q}_2)$ summing to ${\vec q}$ that descends to a 9d state obeying~\eqref{eq:9dWGC}. This is not obvious, since the factor of 2 in~\eqref{eq:10dspectrum} has become a factor of 1 in~\eqref{eq:9dWGC} due to the matching~\eqref{eq:10dto9d}. The idea is to roughly split $\vec q$ in half, though we must guarantee that both ${\vec q}_1$ and ${\vec q}_2$ lie in the $\UE_8$ charge lattice $\Gamma_8$.

\paragraph{Case 1: ${\vec q} \in \Gamma_e$.} First we suppose that the charge vector $\vec q$ has integer entries. We seek to decompose it into ${\vec q}_1$ and ${\vec q}_2$, which either both have integer entries or both have half-integer entries. Without loss of generality, we can write:
\begin{equation}
{\vec q}_1 = \frac{1}{2} {\vec q} + {\vec \delta}, \quad {\vec q}_2 = \frac{1}{2} {\vec q} - {\vec \delta}.
\end{equation}
Then the mass formula~\eqref{eq:10dspectrum} tells us that this state has
\begin{equation}
m^2 = \frac{1}{\alpha'} \left(|{\vec q}|^2 + 4 |{\vec \delta}|^2 - 4\right),
\end{equation}
in which case the 9d WGC criterion~\eqref{eq:9dWGC} holds whenever
\begin{equation} \label{eq:deltabound}
|{\vec \delta}|^2 \leq 1. 
\end{equation}
Thus we must check whether such a $\vec \delta$ exists with ${\vec q}_{1,2} \in \Gamma_8$. 

For a general ${\vec q} \in \Gamma_e$, the vector $\frac{1}{2} {\vec q}$ has a mix of integer and half-integer entries. Because $\sum_i q_i \in 2 \mathbb{Z}$, the entries in $\frac{1}{2} {\vec q}$ sum to an integer, so there are an even number of integer entries and an even number of half-integer entries. Thus, we can choose a candidate $\vec \delta$ that contains either $0$, $2$, or $4$ entries of $\frac{1}{2}$ and the other entries zero, to make the vectors $\frac{1}{2} {\vec q} \pm {\vec \delta}$ have either all integer entries or all half-integer entries. To be more explicit: if $\frac{1}{2} {\vec q}$ contains mostly integer entries, we add $\frac{1}{2}$ to the half-integer entries to make them integers as well, and if $\frac{1}{2} {\vec q}$ contains mostly half-integer entries, we add $\frac{1}{2}$ to the integer entries to make them half-integer entries as well. (If $\frac{1}{2} {\vec q}$ contains four integer entries and four half-integer entries, we can do either.) Now, we evaluate $\sum_i (q_1)_i = \frac{1}{2} \sum_i q_i + \sum_i \delta_i$. If it is an even integer, then so is $\sum_i (q_2)_i$ (because $\sum_i (q_1)_i + \sum_i (q_2)_i = \sum_i q_i \in 2\mathbb{Z}$), and we are done. If it is an {\em odd} integer, and ${\vec \delta} \neq 0$, then we replace ${\vec \delta}$ by flipping precisely one $+\frac{1}{2}$ entry to a $-\frac{1}{2}$ entry. This does not change the integer or half-integer nature of the entries, but it shifts the sum of entries by one, making it even. On the other hand, if ${\vec \delta} = 0$, then we can replace it with a vector with a single $1$ and all other zero entries: ${\vec \delta} = (1, 0, \ldots, 0)$. Again, this does not change the integer or half-integer nature of the entries, and it changes an odd sum to an even one.

At this point, we have constructed ${\vec q}_{1,2} \in \Gamma_8$, and we have a vector ${\vec \delta}$ whose entries are either:
\begin{itemize}
\item All zero, so that $|{\vec \delta}|^2 = 0$.
\item A single $1$ and all others $0$, so that $|{\vec \delta}|^2 = 1$.
\item Two or four entries each of which are $\pm \frac{1}{2}$, and all others zero, in which case $|{\vec \delta}|^2 \leq 4 \left(\frac{1}{2}\right)^2 = 1$.
\end{itemize}
In all of these cases,~\eqref{eq:deltabound} is satisfied, and we have a WGC-satisfying state of charge $\vec q$ in 9d.

\paragraph{Case 2: ${\vec q} \in \Gamma_o$.} If ${\vec q}$ has half-integer entries, and we decompose it as ${\vec q}_1 + {\vec q}_2$ with ${\vec q}_{1,2} \in \Gamma_8$, then we must take one of ${\vec q}_{1,2}$ to lie in $\Gamma_e$ and the other to lie in $\Gamma_o$. We can write each entry in $\vec q$ as the sum of an even integer and either $+\frac{1}{2}$ or $-\frac{1}{2}$:
\begin{equation}
q_i = 2p_i + \frac{1}{2} \sigma_i, \quad p_i \in \ZZ, \quad \sigma_i \in \{ +1, -1\}.
\end{equation} 
From this we learn that 
\begin{equation} \label{eq:qsquaredcase2}
|{\vec q}|^2 = 4 |{\vec p}|^2 + 2 {\vec p} \cdot {\vec \sigma} + \frac{1}{4} |{\vec \sigma}|^2 = 4 |{\vec p}|^2 + 2 {\vec p} \cdot {\vec \sigma} + 2.
\end{equation}
With this decomposition, we define
\begin{equation}
{\vec q}_1 = {\vec p} + {\vec \delta}, \quad {\vec q}_2 = {\vec p} - {\vec \delta} + \frac{1}{2} {\vec \sigma}.
\end{equation}
By construction, these satisfy~\eqref{eq:chargeadd}. From the mass formula~\eqref{eq:10dspectrum} a state with these charges has
\begin{equation} \label{eq:masssquarecase2}
m^2 = \frac{1}{\alpha'} \left(4 |{\vec p}|^2 + 4 |{\vec \delta}|^2 + 2 {\vec p} \cdot {\vec \sigma} - 2 {\vec \delta} \cdot {\vec \sigma}\right),
\end{equation}
and the WGC criterion~\eqref{eq:9dWGC}, comparing~\eqref{eq:masssquarecase2} and~\eqref{eq:qsquaredcase2}, becomes 
\begin{equation}
4 |{\vec \delta}|^2  - 2 {\vec \delta} \cdot {\vec \sigma} \leq 2.
\end{equation}
Now, suppose that $\sum_i p_i \in 2 \mathbb{Z}$. Then we can set ${\vec \delta} = 0$, because ${\vec q}_1 \in \Gamma_e$ and ${\vec q}_2 \in \Gamma_o$. On the other hand, if $\sum_i p_i$ is an odd integer, then we must choose a nontrivial $\vec \delta$. We can take it to be simply ${\vec \delta} = (\sigma_1, 0, 0, \ldots 0)$. This preserves the integer or half-integer nature of any vector it is added to, and shifts the sum of entries by $\pm 1$, so that $\sum_i (q_1)_i$ and $\sum_i (q_2)_i$ become even. In this case, we have 
\begin{equation}
4 |{\vec \delta}|^2  - 2 {\vec \delta} \cdot {\vec \sigma} = 4 \cdot 1 - 2 \cdot \sigma_1^2 = 2.
\end{equation} 

In summary, for any $\vec q \in \Gamma_8$, we are able to write ${\vec q} = {\vec q}_1 + {\vec q}_2$ with ${\vec q}_{1,2} \in \Gamma_8$ such that there is a state of charge $({\vec q}_1, {\vec q}_2)$ in the 10d theory that descends to a WGC-satisfying state of charge $\vec q$ in the 9d theory. This shows that the LWGC holds in the 9d theory, but marginally so, because there are cases in which the 9d WGC inequality was saturated. In other words, the constant $-\frac{4}{\alpha'}$ offset in the mass formula~\eqref{eq:10dspectrum}, which allowed the 10d theory to satisfy the LWGC with apparent room to spare, is {\em precisely} what is required for the 9d theory with a $\ZZ_2$ Wilson line to continue to satisfy the LWGC. Indeed, since a smaller value of the offset would lead to LWGC violation and a larger value would lead to tachyons, we conclude that the offset is uniquely fixed by the joint requirement of satisfying the LWGC with a tachyon-free spectrum.

In what follows, we will consider toroidal orbifold compactifications of heterotic string theory that \emph{do} violate the LWGC. We will see that this violation can be associated with a nontrivial discrete subgroup of the center of the gauge group. In the 9d theory discussed heretofore, the gauge group $E_8$ has a trivial center; it is therefore unsurprising that the LWGC is satisfied in this theory.

\section{Example 2: A $T^4/
\mathbb{Z}_3$ Orbifold}\label{s.T4Z3ex}

We next consider an example of LWGC violation, which was previously discussed in \cite{Heidenreich:2016aqi}. Our treatment of heterotic orbifolds in this section and the following one closely follows that of \cite{Ibanez:2012zz, BHorbifold}.

The theory in question is a $T^4/
\mathbb{Z}_3$ orbifold of the $E_8 \times E_8$ heterotic string, where the $\mathbb{Z}_3$ acts on $T^4$ via
\be
\theta: z_1 \rightarrow e^{2 \pi i/3}z_1\,,~~~z_2 \rightarrow e^{-2 \pi i/3}z_2\,,~~~\Gamma_{16} \rightarrow \Gamma_{16} + \vec{e}_0\,,
\label{orbtwist}
\ee
where
\be
\vec e_0 = \frac{1}{3} (1, -1, 0, ..., 0)\,,
\ee
and each $T^2$ has complex structure $\tau = e^{2 \pi i/3}$.

There is also a Wilson line wrapped around the first circle,
\be
\vec e_1 = \frac{1}{3} ( 0, 0, 1, 1, 2, 0 , 0 , 0)\,.
\ee
For consistency with the orbifold twist, this same Wilson line must also be wrapped around the second circle, as the two circles are rotated into each other under $z_1 \rightarrow e^{2 \pi i/3}z_1$. For simplicity, we assume that all four circles of the torus have the same radius $R$; this assumption will not qualitatively affect our conclusions.

Recall that $\Gamma_{16}$ consists of the product of two $E_8$ lattices. The $E_8$ lattice $\Gamma_{8}$ consists of 8-vectors $(v_1, ..., v_8)$ with with $\sum_i v_i \in 2\mathbb{Z}$ and either
all $v_i \in \mathbb{Z}$ or all $v_i \in \frac{1}{2} + \mathbb{Z}$.

\subsection{Untwisted sector}

It turns out to be sufficient for our purposes to focus our attention on the untwisted sector (indeed, we may also neglect winding and KK momentum around all all cycles of $T^4$ except the ones with the Wilson line). States in the untwisted sector have
\be
\frac{m^2 \alpha'}{4} = \frac{1}{2}(\vec{q}^2 + \vec p_L^2) + N_L - 1 = \frac{1}{2}(\vec p_R^2 + \vec r^2) + N_R - \frac{1}{2}\,.
\label{massformula}
\ee
Here,
\be
\vec{q} = \vec{q}_0 + w \vec{e}_1\,,
\ee
where $w$ is the winding number around the first circle, and the left/right-moving momentum around the first circle is given by
\be
p_{L,R} = \sqrt{\frac{\alpha'}{2}}\left( \frac{1}{R}(k - \vec q_0 \cdot \vec e_1 - \frac{1}{2} \vec e_1^2 w ) \pm \frac{R}{\alpha'} w \right)\,,
\label{pLReq}
\ee
where $k$ is the KK momentum around the first circle. The 4-vector $\vec r$ lies in the $SO(8)$ lattice and controls the spin quantum numbers of the particle in the 10d spacetime. More precisely, $\vec r$ consists of either integers or half-integers, with $\sum_{i=1}^4 r_i$ an odd integer.

The orbifold also imposes a projection condition:
\be
\vec q_0 \cdot \vec e_0 - \vec r \cdot \vec \phi \in \mathbb{Z}\,,
\label{projcond}
\ee
where $\vec \phi = \frac{1}{3}(1, -1, 0, 0)$ comes from the orbifold twist \eqref{orbtwist}.

\subsection{Twisted sectors}

States in the $\theta$-twisted sector have 
\be
\vec{q} = \vec{q}_0 + \vec{e}_0\,,~~~ \vec r = \vec r_0 + \vec \phi\,,
\ee
where $\vec \phi = \frac{1}{3}(1,-1,0,0)$.
These must satisfy the orbifold projection condition
\be
\vec{q} \cdot \vec{e}_0 - \hat r \cdot \vec \phi = \frac{1}{2}(\vec{e}_0^2 - \vec{\phi}^2) \text{ mod }\mathbb{Z}\,,
\label{twistprojcond}
\ee
where $\hat r = \vec r + n + \tilde n - \bar n - \tilde{\bar n}$ includes the effects of oscillators. Here, we conveniently have $\frac{1}{2}(\vec{e}_0^2 - \vec{\phi}^2) = 0$, so the left-hand side of \eqref{twistprojcond} vanishes.

In a theory with a Wilson lines, there are also twisted sectors associated with both a twist $\theta$ and a shift $X^1 \rightarrow X^1 + 2 \pi k R$, $k \in \mathbb{Z}$. String states associated with the fixed point of this roto-translation have gauge momentum
\be
\vec{q} = \vec{q}_0 + \vec{e}_0 + k \vec{e}_1\,.
\ee
These must again satisfy \eqref{twistprojcond}.

\subsection{Gauge group}\label{s.gaugegroup}

Gauge bosons have $\vec r = \pm ( 0, 0 , \underline{0,  1})$ and are massless.\footnote{Here and henceforth, an underline indicates all possible permutations of the numbers above.} By \eqref{massformula}, this means that $\vec p_L =\vec p_R = N_R = 0$. This means that gauge bosons have no winding around the $T^4$, no right-moving oscillators, and they must satisfy
\be
\vec{q}_0 \cdot \vec{e}_1 \in \mathbb{Z} \,,
\label{masslesscond}
\ee
ensuring that the momentum shift $\vec{q}_0 \cdot \vec{e}_1$ in \eqref{pLReq} can be canceled by an appropriate choice of $k$.

The orbifold leaves the second $E_8$ factor unbroken, so it suffices to focus our attention on the first $E_8$ factor.

Eight gauge bosons come from left-moving oscillators (i.e., states with $N_L=1$): these are associated with the Cartan generators. The remaining gauge bosons are associated with the roots of the unbroken gauge group. One can check that the following 36 roots satisfy the orbifold projection condition \eqref{projcond} and the masslessness condition \eqref{masslesscond}:
\begin{align}
\vec{q} \in \{& \pm (1, 1, 0,0,0,0,0,0)\,,~ \pm (0, 0, 1,-1,0,0,0,0)\,,~  \pm (0, 0, \underline{1, 0}, 1, 0, 0, 0) \,, \nonumber \\
&\pm (0, 0, 0,0, 0, \underline{1, \pm 1, 0})\,,~ \pm \frac{1}{2} (+,+, \pm (+,+,-), \pm, \pm, \eta) \}\,.
\end{align}
Here, each $\pm$ is independent, and $\eta = \pm$ is determined uniquely by the requirement that the entries of $\vec{q}$ sum to an even integer.

We next divide these roots into positive and negative roots. Without loss of generality, we choose our positive roots to be those whose first nonzero entry is positive. That is, for a given positive root $\vec{\alpha}=(v_1, ..., v_8)$, there exists $k$ such $v_k > 0$, $v_i = 0$ for $i< k$.

From here, we determine the simple roots, which are the positive roots that cannot be written as positive linear combinations of other positive roots. There are 7 such roots:
\begin{align}
\vec{\alpha}_1 &= \frac{1}{2}(+,+,-,-,+,-,-,+)\nonumber \\
\vec{\alpha}_2 &= (0,0,0,0,0,0,1,-1) \nonumber\\
\vec{\alpha}_3 &= (0,0,0,0,0,1,-1,0) \nonumber \\
\vec{\alpha}_4 &= (0,0,0,0,0,0,1,1)  \label{simpleroots} \\
\vec{\alpha}_5 &=\frac{1}{2}(+,+,+,+,-,-,-,-)\nonumber \\
\vec{\alpha}_6 &= (0,0,1,-1,0,0,0,0)  \nonumber \\
\vec{\alpha}_7 &= (0,0,0,1,1,0,0,0)  \nonumber \,.
\end{align}
One can easily check that the first five roots $\{\vec{\alpha}_1,...,\vec{\alpha}_5\}$ satisfy the relations of the $\mathfrak{su}(6)$ Lie algebra:
\be
\vec{\alpha}_i \cdot \vec{\alpha}_{j} = 2 \delta_{ij} - \delta_{i,j-1}-\delta_{i,j+1}\,.
\ee
Similarly, $\vec{\alpha}_6$ and $\vec{\alpha}_7$ satisfy the $\mathfrak{su}(3)$ relations:
\be
\vec{\alpha}_6 \cdot \vec{\alpha}_6 = \vec{\alpha}_7 \cdot \vec{\alpha}_7 = 2\,,~~\vec{\alpha}_6 \cdot \vec{\alpha}_7 = -1\,.
\ee
Furthermore, $\vec{\alpha}_i \cdot \vec{\alpha}_j = 0$ for $i \leq 5$ and $j \geq 6$.

Thus, the nonabelian part of the gauge algebra is $\mathfrak{su}(6) \oplus \mathfrak{su}(3)$. Since the rank of the gauge group is preserved by the orbifold, we see that there must also be a single $\mathfrak{u}(1)$ factor, so the full gauge algebra is $\mathfrak{su}(6) \oplus \mathfrak{su}(3) \oplus \mathfrak{u}(1)$. To determine the global structure of the gauge group, we compute the spectrum of charged particles.

We begin with the $\mathfrak{su}(6)$ factor. The highest weight of the fundamental representation of $\mathfrak{su}(6)$ is given by the sum of simple roots:
\be
\vec w^1 = A^{1j} \vec{\alpha}_j = \frac{5}{6} \vec{\alpha}_1 + \frac{2}{3} \vec{\alpha}_2 + \frac{1}{2}\vec{\alpha}_3+ \frac{1}{3} \vec{\alpha}_4 + \frac{1}{6} \vec{\alpha}_5\,,
\ee
where here, $A^{ij}$ is the inverse of the Cartan matrix of $\mathfrak{su}(6)$. In the basis used in \eqref{simpleroots}, we have
\be
\vec w^1 = \frac{1}{6}(3,3,-2,-2,2,0,0,0)\,.
\ee
This charge is not in the spectrum: no states in the charge lattice have half-integral $v_1$ and integral $v_8$. In contrast, 
\be
2 \vec w^1 = (1,1,-\frac{2}{3},-\frac{2}{3},\frac{2}{3},0,0,0)\,,
\ee
\emph{does} lie in the spectrum. We conclude that the gauge group must include a $\mathbb{Z}_2$ quotient that acts on the $SU(6)$ factor (and possibly other factors), projecting the fundamental of $SU(6)$ out of the spectrum while leaving the representation with highest weight $2 w^1$.

Meanwhile, the highest weight of the fundamental of $\mathfrak{su}(3)$ is given by
\be
\tilde{w}^1 = \frac{2}{3}\vec{\alpha}_6+ \frac{1}{3} \vec{\alpha}_7 = \frac{1}{3}(0,0,2,-1,1,0,0,0)\,.
\label{su3hw}
\ee
This charge is also in the spectrum: it occurs, for instance, for a winding state with $\vec{q}_0 = (0,0,0,-1,-1,0,0,0)$ and winding number $w=2$ around the first circle, since $\vec{w}^6 = \vec{q}_0 + 2 \vec{e}_1$ for this choice of $\vec{q}_0$. Thus, we learn that the gauge group quotient does not act on the $SU(3)$ factor: the fundamental of $\mathfrak{su}(3)$ resides in the spectrum. We will see below that these fundamentals violate the LWGC.

Finally, the $U(1)$ direction is given by $(1, -1, 0,0,0,0,0,0)$. One can verify that the $\theta$-twisted sector contains states of charges
\be
\vec{q} = \frac{1}{3}(1,-1,0,0,0,0,0,0)\,,~~\vec r=\frac{1}{3}(1,-1,0,3)
\ee
and
\be
\vec{q} = \frac{1}{6}(-1,1,3,3,-3,3,3,3)\,,~~ \vec r=\frac{1}{6}(-1,1,3,3)\,,
\ee
both of which satisfy the projection condition \eqref{twistprojcond}. The first of these carries charge $2$ under $U(1)$, while the second carries charge $-1$ under $U(1)$ and lies in the $\Lambda^3$ of $SU(6)$. Thus, we conclude that the gauge group is
\be
SU(3) \times \frac{SU(6) \times U(1)}{\mathbb{Z}_2}\,.
\ee

\subsection{LWGC violation}

For a given electric charge with winding number $w \neq 0 \mod 3$, there exists values of the radius $R$ such that the LWGC is violated for this charge.

For concreteness, consider the $SU(3)$ charge $\vec{q} =  \frac{1}{3}(0,0,2,-1,1,0,0,0) $ from \eqref{su3hw} above. This charge has winding number $w=2$ with $\vec{q}_0 = (0,0,0,-1,-1,0,0,0)$. From \eqref{pLReq}, this gives
\be
p_{L,R} = \sqrt{\frac{{\alpha}'}{2}}\left( \frac{1}{R}(k +1 - \frac{2}{3} ) \pm \frac{2 R}{{\alpha}'}  \right)\,.
\label{pLRviolation}
\ee
For $R \gg \sqrt{{\alpha}'}$, one can adjust $k$ so that the $1/R$ term essentially cancels out the $R$ term, ensuring that $p_L \lesssim 1$. However, this tuning simultaneously sets $|p_R| \approx 2 \cdot \sqrt{\frac{\alpha'}{2}} \frac{2 R}{\alpha'}  \gg 1$. For $R \gg \sqrt{\alpha'}$, either $p_L$ or $p_R$ (or both) grow parametrically with $R$. By \eqref{massformula}, this means that the mass grows parametrically with $R$ while the charge remains fixed, and the state violates the LWGC. A similar violation occurs also for $R \ll \sqrt{\alpha'}$.

Which states violate the LWGC at large/small radius? From our construction in \S\ref{s.gaugegroup}, we see that nonzero winding is required to produce the fundamental of $\mathfrak{su}(3)$ and the representation of $\mathfrak{su}(6)$ with highest weight $2 \vec w^1$ (which has Dynkin labels $(2, 0,0,0,0)$). Thus, each of these charges violates the LWGC. In addition, a state of charge
\be
\vec{q} = \frac{1}{3}(3, 0, -1, -1, 1 ,0,0,0) 
\ee
necessarily requires winding around the first circle, so it also violates the LWGC: this state lies in the fundamental of $\mathfrak{su}(6)$ and carries charge $3$ under $U(1)$.

In contrast, states with charge $n$, $n \neq 0 \mod 3$ under $U(1)$ necessarily reside in the twisted sector; such states cannot have winding, so they satisfy the WGC.

\begin{table}
\centering
\def\arraystretch{1.5}
\begin{tabular}{|c|cccccc|} \hline
$\mathbb{Z}_6$ charge \textbackslash  $~U(1)$ charge & 0 & 1 & 2 & 3 & 4 & 5  \\ \hline
0 & \textcolor{red}{\Checkmark} & - & \textcolor{red}{\Checkmark}  & - & \textcolor{red}{\Checkmark} & - \\ \hline
1 & - & \Checkmark  & - & \XSolid & - & \Checkmark \\ \hline\
2 & \XSolid & - & \Checkmark & - & \Checkmark & - \\ \hline
3 & - & \textcolor{red}{\Checkmark}  & - & \textcolor{red}{\Checkmark}  & - & \textcolor{red}{\Checkmark}  \\ \hline
4 & \XSolid & - & \Checkmark & - & \Checkmark & - \\ \hline
5 & - & \Checkmark & - & \XSolid & - & \Checkmark \\ \hline
\end{tabular}
\caption{LWGC violation in the $(SU(6)\times U(1))/\mathbb{Z}_2$ sector. Given a charge (mod 6) under the center $\mathbb{Z}_6 \subseteq SU(6)$ and the $U(1)$ factor, \Checkmark indicates the presence of a superextremal state, \XSolid \, indicates a violation of the LWGC, and - indicates that no state exists in the spectrum for this charge. A red \textcolor{red}{\Checkmark} indicates a site in the largest sLWGC-satisfying sublattice. All charges with $n_{U(1)} \neq 0 \mod 3$ have superextremal states because they lie in the twisted sector, and thus they do not involve winding. In contrast, states marked by \XSolid \, require nonzero winding around the first circle, hence they can be made parametrically heavy by increasing $R$.}
\label{tab.LWGCviolation}
\end{table}

The sites in the $(SU(6)\times U(1))/\mathbb{Z}_2$ charge lattice that violate/do not violate the WGC are shown in Table \ref{tab.LWGCviolation}. From this, we conclude that the largest sublattice of $(SU(6) \times U(1))/\mathbb{Z}_2$ that satisfies the sLWGC is the weight lattice of $(\frac{SU(6)}{\mathbb{Z}_3}\times U(1))/\mathbb{Z}_2$.

Finally, since sites with $U(1)$ charge $n \neq 0 \mod 3$ have superextremal states, we also note that the bifundamentals $(\textbf{6}, \overline{\textbf{3}}, \pm 1)$, $(\overline{\textbf{6}}, {\textbf{3}}, \pm 1)$ are superextremal. Similarly, one can check that states of $(p, q)$ under the $\mathbb{Z}_6 \times \mathbb{Z}_3$ center of $SU(6) \times SU(3)$ do not involve winding and hence are superextremal provided that $p+q = 0 \mod 3$. For instance, we have
\be
2 \vec w^1 + 2 \vec{w}^6 = (1, 1, -2, 0,0,0,0,0)\,,
\ee
which lies in the untwisted sector.
Thus, we may expand the superextremal sublattice $\Gamma_{\rm ext}$ to include these states, and we conclude that the maximal superextremal sublattice $\Gamma_{\rm ext}$ at large/small values of the radius $R$ is the weight lattice of
\be
G_{\rm ext} = \frac{\frac{SU(3) \times SU(6)}{\mathbb{Z}_3} \times U(1)}{\mathbb{Z}_2}\,,
\label{gammaext}
\ee
where here the $\mathbb{Z}_2$ acts only on the $SU(6)$ and $U(1)$ factors.

\subsection{Turning off the Wilson line}

Let us next consider the same orbifold but without the Wilson line turned on. From \cite{Etheredge:2025rkn}, we expect that monopoles of the theory without the Wilson line will become (possibly fractionally charged and confined) monopoles of the theory with the Wilson line.

The gauge group of the theory without the Wilson line is larger because there is no projection condition \eqref{masslesscond} on the charges of the W-bosons. The following roots satisfy the orbifold projection condition \eqref{projcond}:
\begin{align}
\vec{q} \in \{& \pm (1, 1, 0,0,0,0,0,0)\,, \pm (0, 0, \underline{1, \pm 1, 0, 0, 0, 0}) \,, \nonumber \\
&\pm (0, 0, 0,0, 0, \underline{1, \pm 1, 0})\,,~ \pm \frac{1}{2} (+, +, \pm, \pm, \pm, \pm, \pm, \eta) \}\,,
\end{align}
where $\eta = \pm 1$ is fixed by the requirement that there are an even number of $-$ signs in total. Altogether, this gives 126 roots plus 8 Cartan generators for a total of 134 gauge bosons, with corresponding gauge algebra $\mathfrak{e}_7 \oplus \mathfrak{u}(1)$.

The simple roots of $\mathfrak{e}_7$ in the chosen basis are given by the following:
\begin{align}
\vec{\alpha}_1 = (0,0,0,0,0,0,1,-1)\,,~~~\vec{\alpha}_2 = (0,0,0,0&,0, 1, -1, 0)\,,~~~\vec{\alpha}_3 = \frac{1}{2}(+, +, -, -, -, -, -, -)\,, \nonumber \\
\vec{\alpha}_4=(0,0,0,0,0,0,1,1)\,,~~~&\vec{\alpha}_5=(0,0,0,0,1,-1,0,0)\,,\\
\vec{\alpha}_6=(0,0,0,1,-1,0,0,0)\,,~~~
&\vec{\alpha}_7=(0,0,1,-1,0,0,0,0)\,.\nonumber
\end{align}
One can verify that their dot products give the $\mathfrak{e}_7$ Cartan matrix; $\vec{\alpha}_2$ corresponds to the trivalent junction in the Dynkin diagram, while $\vec{\alpha}_1$ represents the lone node on top of the trivalent node. Furthermore, the generator of $U(1)$ is given by
\be
\vec{\alpha}_8 = \frac{1}{6}(1, -1, 0,0,0,0,0,0)\,.
\ee

Consider a $\theta$-twisted sector state of charge
\be
\vec{q} = \frac{1}{6}(-1,1,3,3,-3,3,3,3)\,,~~ \vec r=\frac{1}{6}(-1,1,3,3)\,.
\ee
This state represents part of the fundamental of $E_7$ and has charge $-1$ under $U(1)$. In contrast, there are no states in the fundamental of $E_7$, and there and there are no states of $U(1)$ charge 1 in the trivial representation of $E_7$, but there are twisted states of charge $2 \vec{\alpha}_8$, i.e., charge 2 under the $U(1)$.
Thus, we conclude that the global form of the gauge group is $(E_7 \times U(1)) / \mathbb{Z}_2$.

\subsection{Confined monopoles}

Let us denote by $\Gamma$ ($\tilde \Gamma$) the electric (magnetic) charge lattice of the $SU(3) \times (SU(6) \times U(1))/\mathbb{Z}_2$ theory with the Wilson line. We then denote by $\Gamma_0$ ($\tilde \Gamma_0$) the electric (magnetic) charge lattice of the $(E_7 \times U(1))/\mathbb{Z}_2$ theory without the Wilson line. Since $\Gamma$ has additional charges not present in $\Gamma_0$, we have $\Gamma_0 \subsetneq \Gamma$, and thus $\tilde \Gamma \subsetneq \tilde \Gamma_0$.

From \cite{Etheredge:2025rkn}, we expect that the
sLWGC-satisfying sublattice with superextremal particles $\Gamma_{\rm ext}$ should contain the dual of the  lattice including confined monopole charges $\tilde \Gamma_{\rm con}$. Thus we conjecturally have:
\be
\text{Conjecture:}~~~~\Gamma_{\rm ext} \supseteq \tilde \Gamma_{\rm con}^\vee\,.
\label{dualityconj}
\ee

By the argument reviewed above in \S\ref{ss.confinedmonopoles}, confined monopoles are associated with monopoles of the theory without the Wilson line that are not monopoles of the theory with the Wilson line. Hence, by this argument, we have 
\be
\tilde \Gamma_{\rm con} \supseteq \tilde \Gamma_0\,. 
\label{confined0}
\ee

In what follows, we will show that
\be
\Gamma_{\rm ext} \supseteq \Gamma_0\,,
\label{extcontains0}
\ee
with equality occurring at very large (or very small) values of the radius $R$. Combining \eqref{confined0} and \eqref{extcontains0} and using the Dirac quantization relation $\Gamma_0 = \tilde \Gamma_0^\vee$, we find
\be
\Gamma_{\rm ext} \supseteq \Gamma_0 = \tilde \Gamma_0^\vee \supseteq \tilde \Gamma_{\rm con}^\vee\,, 
\label{containchain}
\ee
which establishes the conjecture \eqref{dualityconj}.

To show this, we note that violations of the sLWGC at large or small radius come from untwisted sector states with winding number $w \neq 0 \mod 3$ around the $z_1$ torus. The charges of these states are given by
\be
\vec{q} = \vec{q}_0 + w \vec{e}_1\,,~~~\vec{q}_0 \in \Gamma_{8}\,.
\ee
The shift $w \vec{e}_1$ manifestly requires the presence of the Wilson line.
Thus, these charges do not appear in the lattice $\Gamma_0$ of the theory without the Wilson line; there is no LWGC violation in the lattice $\Gamma_0$, so $\Gamma_{\rm ext} \supseteq \Gamma_0$, as claimed. By \eqref{containchain}, this establishes the desired relation \eqref{dualityconj}.

This result merits further discussion. To begin, note that for intermediate values of the radius, $R \approx \sqrt{\alpha'}$, the LWGC is satisfied, so $\Gamma_{\rm ext} = \Gamma \supsetneq \Gamma_0$, and the containment relation in \eqref{extcontains0} is a proper containment. However, for large or small values of the radius, $R \gg \sqrt{\alpha'}$ or $R \ll \sqrt{\alpha'}$, \eqref{extcontains0} is saturated, and the superextremal sublattice $\Gamma_{\rm ext}$ is precisely equal to the charge lattice of the theory without the Wilson line, $\Gamma_0$.

To see this, recall that for large/small values of the radius $R$, the $\Gamma_{\rm ext}$ is given by the charge/weight lattice of the quotient group $G_{\rm ext}$ shown in \eqref{gammaext}. Meanwhile, the lattice $\Gamma_0$ is the charge/weight  lattice of the group $G_0 = (E_7 \times U(1))/\mathbb{Z}_2$.  
Decomposing the fundamental of $\mathfrak{e}_7$ in terms of its $\mathfrak{su}(6) \oplus \mathfrak{su}(3)$ subalgebra, we have:\footnote{Note that our conventions here differ from those of \texttt{LieART} \cite{Feger:2012bs, Feger:2019tvk} by a complex conjugate on the $SU(3)$ factor: $\textbf{3} \leftrightarrow \overline{\textbf{3}}$.}
\be
\textbf{56} \rightarrow (\textbf{6},\overline{\textbf{3}})+(\overline{\textbf{6}},\textbf{3})+(\textbf{20},\textbf{1})\,.
\ee
Since $(\textbf{56}, \pm 1)$ resides in the charge lattice $\Gamma_0$ of $(E_7 \times U(1))/\mathbb{Z}_2$, we conclude that $(\textbf{6},\overline{\textbf{3}}, \pm 1)$ and $(\overline{\textbf{6}},\textbf{3}, \pm 1)$ also lie in this lattice. Similarly, $(\textbf{1}, \textbf{1} , \pm 2)$ descends from $(\textbf{1}, \pm 2)$, so this charge lies in $\Gamma_0$. Together, these representations of $SU(3) \times (SU(6) \times U(1))/\mathbb{Z}_2$ generate the charge/weight lattice of
\be
G_{\rm ext} = \frac{SU(3) \times \frac{SU(6) \times U(1)}{\mathbb{Z}_2}}{\mathbb{Z}_3}\,,
\label{dualext}
\ee
which is precisely $\Gamma_{\rm ext}$, as determined above in \eqref{gammaext}. This establishes the desired result: for large/small values of the radius $R$, the sLWGC-satisfying sublattice $\Gamma_{\rm ext}$ is identified with the electric charge lattice $\Gamma_{\rm 0}$ of the theory without the Wilson line.

Second, recall that any sublattice of $\Gamma_{\rm ext}$ is itself a superextremal sublattice. It is remarkable, therefore, that the confined monopoles of $\tilde \Gamma_0$ have not merely picked out \emph{a} superextremal, sLWGC-satisfying sublattice, but rather they have distinguished the \emph{maximal} sublattice of particles that remain superextremal for all values of the radius $R$. Consequently, it is consistent with \eqref{dualityconj} to assume that the only confined monopoles in the theory are those which come from the theory without the Wilson line, i.e., $\tilde \Gamma_{\rm con} = \tilde \Gamma_0$. If so, then for sufficiently large or small values of the radius $R$, the conjectured relation \eqref{dualityconj} is in fact an equality: $\Gamma_{\rm ext} = \tilde \Gamma_{\rm con}^\vee$.

Third, it is easy to see that the result \eqref{containchain} generalizes to more general heterotic toroidal orbifolds with Wilson lines and standard embeddings.\footnote{This relation is also satisfied in the simple case of a $\mathbb{Z}_p$ gauge theory compactified on a circle with a Wilson line (cf. \S2 of \cite{Etheredge:2025rkn}).} LWGC violation in such orbifold theories comes from the states in the untwisted sector with nontrivial winding around the torus carrying the Wilson line. Upon turning off the Wilson line, these charges disappear from the spectrum, leaving a sparser lattice $\Gamma_0$ of superextremal states and a finer dual monopole lattice $\tilde \Gamma_0=\Gamma_0^\vee$. By the argument of \S\ref{ss.confinedmonopoles}, monopoles in $\tilde \Gamma_0 \setminus \tilde \Gamma$ become fractionally charged, confined monopoles in the theory with the Wilson line.

Fourth, it is intriguing that even in the presence of LWGC violation, the superextremal sublattice $\Gamma_{\rm ext}$ represents the charge/weight lattice of a Lie group, namely $G_{\rm ext}$ given in \eqref{dualext}. This is the centralizer of the Wilson line element $\vec{e}_1$ in $G_0$, 
and it takes the form
\be
G_{\rm ext} = G/K\,,
\label{Gext}
\ee
where $G = SU(3) \times (SU(6) \times U(1))/\mathbb{Z}_2$ and $K = \mathbb{Z}_3$ is a discrete subgroup of the center of $G$. The coarseness of $\Gamma_{\rm ext}$ follows immediately as the order of the group $\mathbb{Z}_3$. More generally, we expect that $\Gamma_{\rm ext}$ will be the charge/weight lattice of a group $G_{\rm ext}$ of the form given in \eqref{Gext}, with $K \subseteq Z(G)$, and the coarseness of $\Gamma_{\rm ext}$ will be bounded above by $|K|$. This further suggests that if $Z(G)$ is trivial, then the LWGC will be satisfied, which coheres perfectly with the results of \S\ref{s.9dHET}.

Finally, one can explicitly verify the result in  \eqref{dualext} by computing the dot product between the weights $\{\vec{w}_{E_7}^i\}$ of the fundamental representation of the Langlands dual $E_7$ and the lattice sites of $\Gamma$. One finds, for instance, that for all 56 such weights $\vec{w}_{E_7}^i$, the weight of $(\textbf{6}, \overline{\textbf{3}}, -1)$ given by $\vec{q} = \frac{1}{3}(1, 2, -3, 0, 0 , 0 ,0 , 0)$ has dot product
\be
\vec{q} \cdot \vec{w}_{E_7}^i \in \frac{1}{2} + \mathbb{Z}\,,~~~i=1,...,56.
\ee
Meanwhile, the unit charge $U(1)$ monopole has magnetic charge vector
\be
\vec{q}_{\rm mag} = \frac{3}{2}(1, -1, 0, 0 , 0 , 0 , 0 , 0)\,,
\ee
which has $\vec{q} \cdot \vec{q}_{\rm mag} = -\frac{1}{2}$. Thus, any weight of the dual $(E_7 \times U(1))/\mathbb{Z}_2$ has dot product
\be
\vec{q} \cdot (\vec{w}_{E_7}^i+\vec{q}_{\rm mag}) \in \mathbb{Z}\,.
\ee
That is, any confined monopole in $\tilde \Gamma_0$ has integral Dirac pairing with a weight of $(\textbf{6}, \overline{\textbf{3}}, -1)$, which corresponds to the fact that this weight lies in $\Gamma_{\rm ext}$, which is dual to $\tilde \Gamma_0$.

In contrast, letting $\vec{q}=2 w^1 = \frac{1}{3}(3,3,-2,-2,2,0,0,0)$, we find that 
\be
\vec{q} \cdot (\vec{w}_{E_7}^i+\vec{q}_{\rm mag}) \in \frac{1}{3} + \mathbb{Z}
\ee
for some choices of $\vec{w}_{E_7}^i$. This implies that this lattice site is not in the dual of $\tilde \Gamma_{\rm con} = \tilde \Gamma_0$, which coheres perfectly with the fact that $\vec{q}$ is not in the superextremal sublattice $\Gamma_{\rm ext}$ for sufficiently large or small values of the radius.

\subsection{Splitting and flux tube tensions}

All of the examples studied in \cite{Etheredge:2025rkn} 
also displayed a more quantitative relation between LWGC violation and monopole confinement. Let us define the \emph{splitting mass} via
\be
m_{\rm split}^2 \equiv \max \left (0 , \inf_{q \in \Gamma \setminus \Gamma_{\rm ext}} m(q)^2 - m_{\rm ext}(q)^2 \right)\,,
\label{msplit}
\ee
where $m_{\rm ext}(q)$ is the mass of a hypothetical, extremal particle of charge $q$.
Note that $m_{\rm split}$ vanishes in any theory that satisfies the LWGC.

With this definition, the relation observed in \cite{Etheredge:2025rkn} can be stated succinctly as follows: large splitting ($m_{\rm split} \gg 1$ in Planck units) implies light flux tubes ($T_{\rm flux} \ll 1$ in Planck units). This relationship is not so easy to verify here, for several reasons.

To begin, according to the definition introduced in \eqref{msplit}, the splitting mass $m_{\rm split}$ actually vanishes for all values of the parameters $g_s$ and $R$.
The reason is that, as discussed in \S3.2 of \cite{Heidenreich:2016aqi}, LWGC violation in this theory occurs only at a finite number of sites in the charge lattice. Thus, given a fixed radius $R \gg \sqrt{\alpha'}$ or $R \ll \sqrt{\alpha'}$, we may choose our winding $w$ and KK momentum $k$ in \eqref{pLRviolation} so that $p_L^2 \leq 2$, possibly at the expense of $p_R^2 \gg 1$. However, if we restrict our attention to charges $\vec{q}$ with $\vec{q}^2 \gg p_R^2$, we can ensure that the large right-moving mass contribution from $ p_R^2$ does not render the state subextremal. Therefore, sufficiently far out on the charge lattice, all charges have superextremal particles, so $m_{\rm split}$ vanishes.

Indeed, $m_{\rm split}$ vanishes even in the asymptotic limits $R \rightarrow \infty$, $R \rightarrow 0$, since there exist an infinite collection of superextremal, twisted sector states outside $\Gamma_{\rm ext}$; these are denoted with a black \Checkmark in Table \ref{tab.LWGCviolation}. Thus, at least within the controlled regime of the theory with weak heterotic string coupling $g_s < 1$, there is no limit in which the splitting mass $m_{\rm split}$ is super-Planckian (or even nonzero). This conclusion applies more generally to heterotic orbifolds with Wilson lines and standard embeddings.

The magnetic side of the story is more mysterious. As discussed in \S\ref{ss.confinedmonopoles}, confining flux tubes descend from vortices of the higher-dimensional theory, wrapped over the compactification geometry. In the case at hand, these vortices are \emph{Gukov-Witten twist vortices}---dynamical versions of (non-topological) Gukov-Witten operators \cite{Gukov:2008sn}. To our knowledge, the properties of these vortices (and in particular their tensions) have not been studied in the literature.

Nonetheless, we may estimate the dependence of the flux tube tension on the $T^4$ volume, $\text{vol}_{T^4} \sim R^4$, in the large volume regime $R \gg \sqrt{\alpha'}$. We have in 6d: 
\be
T_{\rm flux} \sim \text{vol}_{T^4} T_{\rm GW}\,,
\ee
where $T_{\rm GW}$ is the tension of the Gukov-Witten operator in 10d. Working in 6d Planck units with $M_{\rm Pl; 6} = 1$, we have
\be
T_{\rm flux} \sim  \frac{T_{\rm GW}}{M_{\rm Pl; 10}^8}\,,
\ee
hence the radial dependence has canceled out. We conclude that the flux tube tension does not tend to zero in the limit $R \rightarrow \infty$, which fits with the observation that the splitting does not diverge in this limit.

Thus, in contrast with the examples studied in \cite{Etheredge:2025rkn}, it is unclear if there exists a limit in moduli space where large splitting and light flux tubes occur and induce a dynamical change of the gauge group from $G$ to $G/K$. Answering this question would require a thorough investigation of charged particles and Gukov-Witten vortices in the strongly coupled regime of moduli space with $g_s \gg  1$. This regime likely features warping of the type studied in e.g. \cite{Aharony:2007du, Etheredge:2023odp}. We leave these questions for future study.

One thing that the present examples \emph{do} have in common with those of \cite{Etheredge:2025rkn} is the presence of massive vector bosons. In particular, the W-bosons associated with roots of the gauge group $G_0$ that are not roots of the gauge group $G$ acquire a mass in the presence of the Wilson line. In addition, there are massive KK modes of the (massless) gauge bosons both with and without the Wilson line. In the examples of \cite{Etheredge:2025rkn}, the vector boson mass $m_\gamma$ was bounded below (in Planck units) by the product of the flux tube tension and the splitting mass, $m_{\gamma} \gtrsim m_{\rm split} T_{\rm flux}$. It is fascinating that massive vector bosons represent a hallmark of this class of LWGC-violating heterotic orbifold models even though $m_{\rm split}$ vanishes.

\section{Example 3: Another $T^4/\mathbb{Z}_3$ Orbifold}\label{s.T4Z3ex2}

In this section, we consider another example of LWGC violation, which was previously discussed in \cite{BHorbifold}. This example differs from that of the previous section in that all monopoles in the theory without the Wilson line are unstable. Nonetheless, we will show that at least some of these monopoles can be stabilized by a different choice of Wilson line.

The theory in question comes from another $T^4/\mathbb{Z}_3$ orbifold, where the $\mathbb{Z}_3$ acts on $T^4$ via
\be
\theta: z_1 \rightarrow e^{2 \pi i/3}z_1\,,~~~z_2 \rightarrow e^{-2 \pi i/3}z_2\,,~~~\Gamma_{16} \rightarrow \Gamma_{16} + \vec{e}_0\,,
\ee
with
\begin{equation}
  \vec{e}_0 = \frac{1}{3} (-1,-1,-1,-1,-1,-1,-1,7)  \,.
\end{equation}
Again, both $T^2$ factors have complex structure $\tau = e^{2 \pi i/3}$. 

We further consider the effect of a Wilson line on the first torus:
\be
\vec{e}_1 = \frac{1}{3}(1,1,1,-1,-1,-1,0,0)\,.
\label{ex2WL}
\ee
Before the Wilson line is turned on, the surviving gauge group is $SU(9)$ \cite{BHorbifold}. With the Wilson line turned on, the gauge group is broken to $(SU(3)_1 \times SU(3)_2 \times SU(3)_3 \times U(1)_1 \times U(1)_2)/\mathbb{Z}_6$, and the simple roots are given by
\begin{align}
\vec{\alpha}_1 = (1, -1, 0, 0 ,0 ,0 ,0,0)\,,~~~\vec{\alpha}_2 = (0, 1, -1, 0 ,0 ,0 ,0,0) \,,~~~\vec{\alpha}_3 = (0, 0, 0, 1, -1, 0 ,0 ,0 )  \\ \nonumber
 \vec{\alpha}_4 = (0, 0 ,0, 0, 1, -1 ,0 ,0)\,,~~~\vec{\alpha}_5 = (0, 0, 0, 0 ,0 ,0 ,1,1) \,,~~~\vec{\alpha}_6 =  \frac{1}{2} (+,+,+,+,+,+,-,-)\,.
\end{align}
Here, $(\vec{\alpha}_{2n -1}, \vec{\alpha}_{2n})$ are the simple roots of $SU(3)_n$, $n=1,2,3$. In addition, there are two $U(1)$ factors. The $U(1)$ generators are given by
\be
\vec{\alpha}_7 = \frac{1}{6}(0, 0, 0, 0 , 0, 0 , -1, 1)\,,~~~\vec{\alpha}_8 = -\frac{1}{6}(1,1,1,-1,-1,-1,0,0) = -\vec{e}_1 \,.
\ee
By explict computation, we find that the generators of the $\mathbb{Z}_6$ quotient act on the center $\mathbb{Z}_3^3 \times U(1)^2$ symmetry with the following charges:
\be
\mathbb{Z}_3: (1,1,-1,-1,0)\,,~~~
\mathbb{Z}_2: (0,0,0,1,1)\,.
\ee
The former of these means, for instance, that the fundamental $\textbf{3}_i$ of $SU(3)_i$ does not appear in the spectrum. However, the bifundamentals $(\textbf{3}, \overline{\textbf{3}}, \textbf{1},0,0)$, $(\textbf{3}, \textbf{1}, \textbf{3},0,0)$, and  $(\textbf{1}, \textbf{3}, \textbf{3},0,0)$ do appear in the spectrum. The latter quotient implies that the sum of the two U(1) charges must be even.

Meanwhile, under the branching $\mathfrak{su}(9) \rightarrow \mathfrak{su}(3)^3 \oplus \mathfrak{u}(1)^2$, the weight lattice of $SU(9)$ descends to the weight lattice of
\be
(SU(3)_1 \times SU(3)_2 \times SU(3)_3 \times U(1)_1 \times U(1)_2)/(\mathbb{Z}_6 \times \mathbb{Z}_3)\,.
\ee
This additional $\mathbb{Z}_3$ quotient implies that $\Gamma \supsetneq \Gamma_0 = \Gamma_{\rm ext}$: the superextremal sublattice is a proper sublattice of the electric charge lattice in the theory with the Wilson line. LWGC violation once again comes from untwisted sector states with winding $w \neq 0 \mod 3$ around the torus with the Wilson line.

So far, our analysis of the electric spectrum of this theory closes parallels that of \S\ref{s.T4Z3ex} above. The novelty in this example lies in the magnetic side of the theory. It is still true that $\Gamma_{\rm ext} = \tilde \Gamma_0^\vee$ for sufficiently large or small values of the compactification radius $R$: the sLWGC-satisfying sublattice is dual to the magnetic weight lattice $\tilde \Gamma_0^\vee$ of the $SU(9)$ gauge theory. However, as discussed in \S\ref{ss.NAREVIEW}, since $\pi_1(SU(9))$ is trivial, none of the monopoles are stable.

However, there is a way around this subtlety. In what follows, we will show that by turning on a different Wilson line, we can turn unstable monopoles of the $SU(9)$ into genuine, stable monopoles. These monopoles, in turn, produce fractionally charged confined monopoles in the $SU(3)^3\times U(1)^2$ theory.

\subsection{A different Wilson line}

Suppose that instead of the Wilson line of \eqref{ex2WL}, we turn on the Wilson line
\be
\vec{e}_2 = \frac{1}{3}(0,1,-1,0,0,0,0,0)\,.
\label{ex2WL2}
\ee
on the first torus. This breaks the gauge group to
\be
(SU(7) \times U(1)_1 \times U(1)_2)/(\mathbb{Z}_{7} \times \mathbb{Z}_2)\,,
\ee
where the discrete quotient acts on the $\mathbb{Z}_7 \times U(1)_1 \times U(1)_2$ center symmetry with charges 
\be
\mathbb{Z}_7: (2,1,0)\,,~~~
\mathbb{Z}_2: (0,1,1)\,.
\ee
Here, the simple roots of $\mathfrak{su}(7)$ are given by
\begin{align}
\vec{\alpha}_1 = \frac{1}{2} (+, -, -, - ,- ,- ,-,+)\,,~~~ \vec{\alpha}_2 = \frac{1}{2} (+, +, +, - ,+ ,+ ,+,-) \,,~~~\vec{\alpha}_3 = (0, 0, 0, 1, -1, 0 ,0 ,0 )  \\ \nonumber
\vec{\alpha}_4 = (0, 0 ,0, 0, 1, -1 ,0 ,0)\,,~~~ \vec{\alpha}_5 = (0, 0, 0, 0 ,0 ,1 ,-1,0) \,,~~~ \vec{\alpha}_6 =  (0, 0, 0, 0 ,0 ,0 ,1,1) \,.
\end{align}
Their dot products yield the Cartan matrix of $\mathfrak{su}(7)$: $\vec{\alpha}_i \cdot \vec{\alpha}_j = 2 \delta_{i j} - \delta_{i, j+1} - \delta_{i, j-1}$. In addition, the generators of $U(1)_1 \times U(1)_2$ are given respectively by
\be
\vec{\alpha}_7 = \frac{1}{42}(-1,2,2,-1,-1,-1,-1,1)\,,~~~\vec{\alpha}_8 = \frac{1}{6}(0,1,-1,0,0,0,0,0)\,.
\ee

For the sake of concreteness, let us consider the lightest state with electric charge vector
\begin{align}
\vec{q} = (-1,-1,0,0,0,0,0,0) + \frac{1}{3}(1,1,1,-1,-1,-1,0,0) 
\nonumber \\= \frac{1}{3}(-2,-2,1,-1,-1,-1,0,0)\,,
\end{align}
which lies in the untwisted sector of the $(SU(3)^3 \times U(1)^2)/\mathbb{Z}_6$ theory with Wilson line $\vec{e}_1$ and violates the LWGC. We want to show that there exists a genuine monopole of the $(SU(7) \times U(1)^2)/\mathbb{Z}_{14}$ theory with Wilson line $\vec{e}_2$ that has fractional Dirac pairing with this $\vec{q}$. One example of such a monopole is given by
\be
\vec{q}_{\rm mag} = \frac{1}{2}(-3,5,5,-1,-3,-3,-3,3)\,.
\ee
This vector corresponds to a weight of the $\bar {\bf{7}}$ of the Langlands dual $SU(7)$ and carries magnetic charge $(6,0)$ under $U(1)_1 \times U(1)_2$. This charge is projected in by the $\mathbb{Z}_{14}$ quotient on the magnetic charge lattice, so it is a genuine, topologically stable monopole of the $(SU(7) \times U(1)^2)/\mathbb{Z}_{14}$ theory. 

One can easily check that
\be
\vec{q} \cdot \vec{q}_{\rm mag} = \frac{4}{3}\,,\label{fraceq}
\ee
which shows that this monopole charge $\vec{q}_{\rm mag}$ violates Dirac quantization in the $(SU(3)^3 \times U(1)^2)/\mathbb{Z}_6$ theory. Instead, by the argument of \cite{Etheredge:2025rkn}, we may push this monopole through a domain wall from the $\vec{e}_2$ theory to the $\vec{e}_1$ theory to produce a confined monopole, and \eqref{fraceq} shows that this monopole has fractional Dirac pairing with the LWGC-violating state in question. This confirms our hypothesis: LWGC-violating states have fractional Dirac pairing with confined monopoles.

\section{Discussion}\label{s.DISC}

In this work, we have seen evidence that the connection between LWGC violation and fractionally charged confined monopoles---first uncovered in \cite{Etheredge:2025rkn}---extends to the case of nonabelian gauge groups as well. More precisely, we have seen that in heterotic toroidal orbifolds with discrete Wilson lines, the sublattice of superextremal particles $\Gamma_{\rm ext}$ contains the electric charge lattice of the theory without Wilson lines $\Gamma_0$, with equality obtained at sufficiently large or sufficiently small values of the compactification radius. The dual of this lattice $\tilde \Gamma_0$ contains the lattice of fractionally charged confined monopoles $\tilde \Gamma_{\rm con}$, which implies that the sublattice of superextremal particles contains the dual of the superlattice of monopoles including those that are confined, $\Gamma_{\rm ext} \supseteq \tilde \Gamma_{\rm con}^\vee$.

We have also uncovered a fascinating connection between LWGC violation and the global structure of the gauge group. In all of the examples studied, $\Gamma_{\rm ext}$ represents the charge lattice of a quotient group $G_{\rm ext} = G/K$, where $G$ is the gauge group of the theory and $K$ a discrete subgroup of $Z(G)$.
The coarseness of the sublattice $\Gamma_{\rm ext}$ is given by the maximal order of the elements of $K$. This suggests that, in general, the LWGC will be satisfied if $Z(G)$ is trivial. Indeed, in \S\ref{s.9dHET}, we verified that the LWGC is precisely saturated in a 9d background of the heterotic string theory with gauge group $G = E_8$, $Z(G) = \{1\}$.

In general, in a limit of large splitting $m_{\rm split} \rightarrow \infty$, the subextremal particles outside $\Gamma_{\rm ext}$ exit the spectrum, and the gauge group changes to the quotient group $G_{\rm ext} = G/K$ as an 1-form discrete $K$ symmetry emerges. On the magnetic side, fractionally charged monopoles deconfine and become genuine monopoles. In the heterotic orbifolds studied here, however, we have not encountered any limits of large splitting. It is intriguing that a discrete group $K \subseteq Z(G)$ nonetheless plays a starring role in these models, even though it does not seem to arise as a physical 1-form gauge symmetry.

Finally, our discussion of splitting and flux tubes led us to the fascinating yet unexplored topic of Gukov-Witten vortices: dynamical versions of the celebrated Gukov-Witten operators. These vortices have been discussed under the name of ``twist vortices'' or ``Alice strings'' in theories with disconnected gauge groups \cite{Heidenreich:2021tna}, where their existence follows from the absence of higher-form global symmetries in quantum gravity \cite{Hawking:1974sw, Banks:2010zn, Harlow:2018tng}. However, further analysis of these vortices---particularly in theories with connected gauge groups---would be most fascinating.

Even with the recent progress in our conceptual understanding of the sLWGC, we are still left with a significant open question: how large can the coarseness $k$ be? In this work, we have proposed that $k$ is bounded above by the maximal order of an element of the center of the gauge group. This is a very useful bound for nonabelian gauge theories, but it offers no immediate restriction on the coarseness in $U(1)$ gauge theories. Nonetheless, it would be interesting to see if the restriction on nonabelian gauge groups can be leveraged to produce a universal bound in the abelian case, at least in situations where $U(1)$s arise via adjoint Higgsing. For example, it has been argued that the largest permissible rank of a nonabelian gauge theory is $26 - d$ for gravitational theories with 16 supercharges in $d > 3$ dimensions~\cite{Kim:2019ths}. It is not so clear what can be said with less (or no) supersymmetry.

Our understanding of the WGC has grown significantly in the twenty years since its formulation, and even today it continues to spill its secrets. It is our hope and expectation that another twenty years of exploration will bring new insight into both the WGC and the heart of quantum gravity.

\section*{Acknowledgments}

We thank Muldrow Etheredge and Sebastian Rauch for useful comments on a draft of this paper. MR is supported in part by the DOE Grant DE-SC0013607. The work of TR was supported in part by STFC through grant ST/T000708/1 and by the Royal Society through grant RGS/R2/252603.

\bibliographystyle{utphys}
\bibliography{ref}
\end{document}